\documentclass[aps]{revtex4}
\usepackage[dvips]{graphics,graphicx}
\begin{document}
\title{Damped Bloch oscillations of cold atoms in optical lattices}
\author{A. R. Kolovsky}
\author{A. V. Ponomarev}
\affiliation{Kirensky Institute of Physics, 660036 Krasnoyarsk, Russia}
\author{H. J. Korsch}
\affiliation{Universit{\"a}t Kaiserslautern, FB-Physik, D-67659 Kaiserslautern, Germany}
\date{\today }

\begin{abstract}
The paper studies Bloch oscillations of cold neutral atoms in the
optical lattice. The effect of spontaneous emission on the dynamics
of the system is analyzed both analytically and numerically.
The spontaneous emission is shown to cause (i) the decay of
Bloch oscillations with the decrement given by the rate of spontaneous
emission and (ii) the diffusive spreading of the atoms with a
diffusion coefficient depending on {\em both} the rate of spontaneous
emission and the Bloch frequency.
\end{abstract}

\maketitle

%%%%%%%%%%%%%%%%%%%%%%%%%%%%%%%%%%%%%%%%%%%%%%%%%%%%%%%%%%%%%%%%%
\section{Introduction}

In 1928 Bloch predicted that the coherent motion of crystal
electrons in a static electric field should be oscillatory
rather than uniform \cite{Bloc28}. Nevertheless, this phenomenon
(known nowadays as Bloch oscillations), has never been observed
in bulk crystals. Because of relaxation processes (scattering
on lattice defects, phonons, etc.) the coherence of the system is
destroyed  before electrons complete one Bloch cycle. This
obstacle has forced the researchers to look for other systems,
where the Bloch period (which is inversely proportional to the
magnitude of the static force and the lattice period) can be smaller
than the characteristic relaxation time. This is realized in
semiconductor superlattices, and in 1992 a direct observation
of the current oscillations in semiconductor superlattices was
reported \cite{Feld92,Leo92}. It should be stressed, however,
that in semiconductor superlattices the relaxation time only
slightly exceeds the Bloch period and, thus, in practice one
always meets the regime of {\em damped} Bloch oscillations.
A detailed analysis of the relaxation processes
in the semiconductor system appears to be a rather complicated
problem and phenomenological approaches are usually used to
describe the decay of current oscillations.

Recently Bloch oscillations were observed in the system of cold
atoms in an (accelerated) optical lattice \cite{Daha96}. This
system of atomic optics mimics the solid state system where the
neutral atoms and standing laser wave play the roles of electrons
and crystal lattice, respectively. Unlike the solid state
systems, it may show both the regimes of undamped and damped
oscillations. Indeed, the main relaxation process in the system
``atom in a laser field" is the spontaneous emission of photons
by the excited atom and it can be well controlled by choosing an
appropriate detuning from the atomic resonance. Up to now only
the regime of undamped oscillations (large detuning) has
attracted the attention of researches
\cite{Daha96,Raiz97,Ande98,PLA00,Mors01,Thom02,PR02}.
However, keeping in mind an analogy
with the semiconductor superlattices, the latter regime
is also of considerable interest. In this present paper we study
the process of decay of atomic Bloch oscillations due to the
effect of spontaneous emission. In particular we address
the question of the actual decay process and the dependence of the
decay rate on the system parameters.

%%%%%%%%%%%%%%%%%%%%%%%%%%%%%%%%%%%%%%%%%%%%%%%%%%%%%%%%%%%%%
\section{The model and approach}

We shall take into account the spontaneous emission by using the
standard approach, where its effect is characterized by a single
constant $\gamma$ defined as the inverse radiative lifetime of the
upper state. Namely, it is assumed that the diagonal and
off-diagonal elements of the $2\times2$ density matrix of a fixed
atom relax to their equilibrium values with the rates $\gamma$ and
$\gamma/2$, respectively. When atom is in free space, its density
matrix depends additionally on the position $z$ of the atom and,
in general case, the resulting master equation has a rather complicated form
\cite{Leto81,Adam94}. This equation can be considerably simplified in the
limit of large detuning $\delta\gg\Omega$ (here $\Omega$ and
$\delta$ are Rabi frequency and detuning from the atomic
resonance). In this case we can eliminate the internal structure of
the atom and the master equation for the position density matrix
has the form
%************************************************************
\begin{equation}
\frac{\partial\rho}{\partial t}=-\frac{i}{\hbar}[H,\rho \,]
-\frac{\gamma}{2}\frac{\Omega^2}{\delta^2}\int {\rm d}u P(u)
\left(L^\dagger_u L_u\rho-2L_u\rho L^\dagger_u+\rho L^\dagger_u
L_u\right) \;,
\label{1}
\end{equation}
(see, Ref.~\cite{Grah96,Goet96}, for example) where
%************************************************************
\begin{equation}
H=\frac{p^2}{2M}-\frac{\hbar\Omega^2}{\delta}\cos^2(k_L z)+Fz
\label{2}
\end{equation}
is the effective Hamiltonian of the system and
%************************************************************
\begin{equation}
\label{3}
L_u=\cos(k_L z)\exp(iuk_L z) \;,\quad |u|\le 1
\end{equation}
is the projection of the recoil operator on $z$-axis. The distribution
$P(u)$ of a random variable $u$ in Eq.~(\ref{1}) is defined by the
angle distribution for the momentum $\hbar{\bf k}_L=\hbar k_L{\bf
n}$ of the spontaneously emitted photons, which in the case of
linearly polarized light is given by $\Phi({\bf n})
=(3/8\pi)[1-({\bf n}\cdot{\bf e})^2]$ \cite{Leto81}.
Note that  equation (\ref{1}) has the Lindblad form and,
thus, ${\rm Tr}[\rho(t)]=\int{\rm d}z \rho(z,z,t)=1$. Beside this,
for linearly polarized light, $P(u)$ appears to be almost independent
of $u$ and for most practical purposes one can set $P(u)=1/2$.

To solve Eq.~(\ref{1}), we use a Monte-Carlo method of
Ref.~\cite{Goet96}. According to this method one finds the
dynamics of an arbitrary observable $\langle A(t)\rangle={\rm Tr}
[A\rho(t)]$ as an average
%************************************************************
\begin{equation}
\langle A(t)\rangle=\overline{\langle \psi(t)|A|\psi(t)\rangle}
\label{4}
\end{equation}
over different solutions of the stochastic Schr{\"o}dinger
equation
%************************************************************
\begin{equation}
{\rm d}\psi=\left(-\frac{i}{\hbar}H{\rm d}t
-\frac{\tilde{\gamma}}{2}L^\dagger_u L_u {\rm d}t
+\sqrt{\tilde{\gamma}}L_u{\rm d}\xi \right)\psi \;.
\label{5}
\end{equation}
In the latter equation $\tilde{\gamma}=\gamma(\Omega/\delta)^2$
is the spontaneous emission rate and ${\rm d}\xi$ is a Wiener process
with $\overline{{\rm d}\xi}=0$ and $\overline{{\rm d}\xi^2}={\rm d}t$
(i.e., $\xi(t)$ is $\delta$-correlated white noise). It is worth
to note that the choice of the stochastic Schr{\"o}dinger
equation is not unique and one can construct many different
stochastic equations corresponding to the same master equation.
Most often the nonlinear versions, which preserve the norm of the
wave function $\psi(t)$, are employed \cite{Meystre}. Here,
following Ref.~\cite{Goet96}, we use a linear stochastic equation
which is preferable from the numerical point of view. Then the norm is
conserved only after averaging over different realizations of
the random process $\xi(t)$. This fact was used to control the
statistical convergence.

%%%%%%%%%%%%%%%%%%%%%%%%%%%%%%%%%%%%%%%%%%%%%%%%%%%%%%%%%%%%
\section{Tight-binding approximation}

First we shall analyze the problem in the tight-binding
approximation. In this approximation the Hamiltonian (\ref{2}) is
substituted by the tri-diagonal matrix
%************************************************************
\begin{equation}
\label{6}
H_{l,m}=-\frac{\Delta}{2}(\delta_{l,m+1}+\delta_{l,m-1})
+dFl\delta_{l,m}  \;,
\end{equation}
and the recoil operator (\ref{3}) by the diagonal matrix
%************************************************************
\begin{equation}
\label{7}
L_{l,m}(u)=(-1)^ l\exp(i\pi ul)\delta_{l,m} \;.
\end{equation}
Here $d=\pi/k_L$ is the period of the optical potential and the index $l$
refers to the localized Wannier function $|l\rangle$ associated
with the $l$-th well of the periodic potential. In the absence of
spontaneous emission ($\gamma=0$) and static force ($F=0$)
the eigenfunctions of the system are Bloch waves
$|\psi_\kappa\rangle=\sum_l\exp(id\kappa l)|l\rangle$
corresponding to the energy $\epsilon(\kappa)=-\Delta\cos(d\kappa)$.
This dispersion relation is assumed to approximate the dispersion
relation of the atom in the ground Bloch band. In the presence of
a static force, the quasimomentum $\kappa$ of the wave function
$|\psi_\kappa\rangle$ evolves according to the classical equation
${\rm d}\kappa/{\rm d}t=F/\hbar$. In terms of the atomic
velocity $v=\hbar^{-1}\partial \epsilon(\kappa)/
\partial\kappa$ and coordinate $z$ this corresponds to a
periodic oscillation of the wave packet with the Bloch frequency
$\omega_B=dF/\hbar$:
%************************************************************
\begin{equation}
\label{9}
\langle v(t)\rangle=\frac{d\Delta}{\hbar}\sin(\omega_B t) \;,\quad
\langle z(t)\rangle=\frac{\Delta}{F}\cos(\omega_B t) \;.
\end{equation}
(The coordinate and velocity operators are obviously given by the
matrices $z_{l,m}=dl\delta_{l,m}$ and
$v_{l,m}=(d\Delta/2\hbar)(i\delta_{l,m+1}-i\delta_{l,m-1}$).) Note
that within the tight-binding approximation the atoms always
oscillate according to a cosine law. In reality, however, the
oscillations are asymmetric. Besides this, there is a decay of
the oscillations due to the interband Landau-Zener tunneling
\cite{Land32,Zene34,Bhar97,JOPB00a} -- a
phenomenon completely ignored by the tight-binding (and more
general single-band) models. With these remarks reserved, we
proceed with the analysis of the effect of spontaneous emission.

As follows from the explicit form of Eq.~(\ref{5}), the recoil
operator randomly changes the atomic quasimomentum. Thus any
narrow distribution of the quasimomentum (which is usually
considered as a prerequisite for observing Bloch oscillations)
will be smeared over the entire Brillouin zone and the oscillations
should decay. The following simple ``classical" model helps to
understand the details of the decay process. Let us consider an
ensemble of classical particles with the Hamiltonian
%************************************************************
\begin{equation}
\label{8}
H=-\Delta\cos(dp/\hbar)+Fz
\end{equation}
affected additionally by noise (random recoil kicks) so that the 
conditional probability is %\cite{remark2}
%************************************************************
\begin{equation}
\label{12}
W(p,p',t)=(2\pi\tilde{\gamma}t)^ {-1/2}
\exp\left\{-\frac{[d(p-p'-Ft)/\hbar]^2}{4\tilde{\gamma}t}\right\} \;.
\end{equation}
Then the mean velocity $\langle v(t)\rangle=\langle \partial
H/\partial p\rangle$ decays as
%************************************************************
\begin{equation}
\label{10}
\langle v(t)\rangle=v_0\exp(-\tilde{\gamma}t) \sin(\omega_B t)
\;,\quad v_0=d\Delta/\hbar
\end{equation}
and the mean squared velocity relaxes to $v^2_{st}=v_0^2/2$.

\begin{figure}
\center
\includegraphics[width=12.5cm, clip]{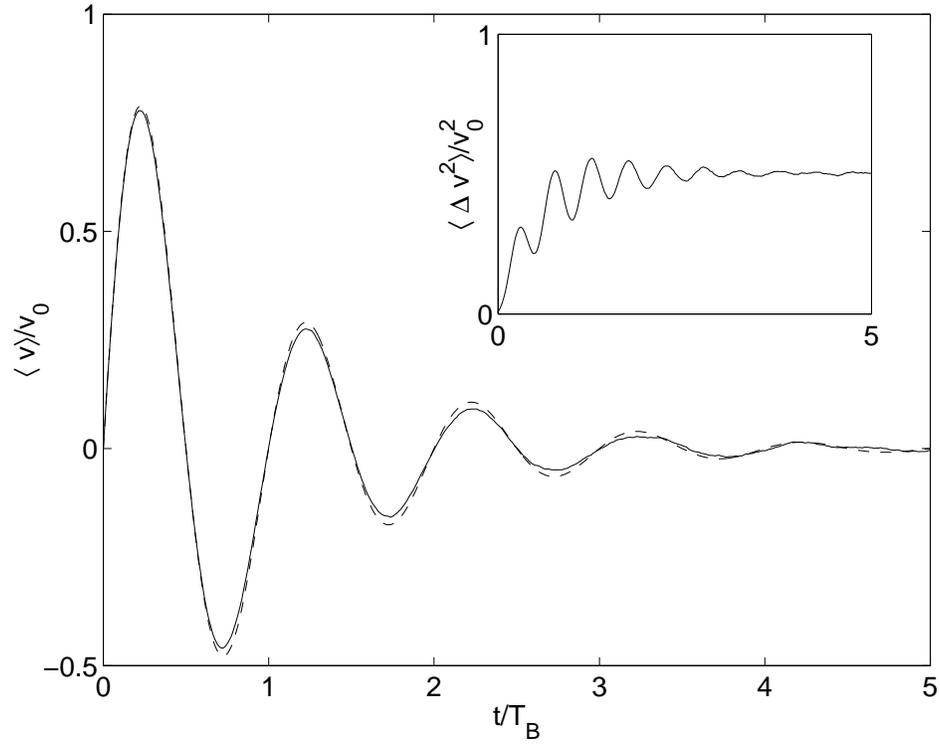}
\caption{Tight-binding model. Dynamics of the mean velocity (solid
line) and the mean squared velocity (inset) for
$\tilde{\gamma}=0.05$ and $F=-0.1$. The dashed line corresponds to
Eq.~(\ref{10}).} \label{fig1}
\end{figure}

The solid line in Fig.~\ref{fig1} shows the dynamics of $\langle
v(t)\rangle$ calculated on the basis of equations
(\ref{5}) -- (\ref{7}), where as an initial condition we choose
$\rho(0)=|\psi(0)\rangle\langle\psi(0)|$ with $\psi(0)\sim
\sum_l\exp(-l^2/100)|l\rangle$ and set $\Delta=1$, $\hbar=1$,
and $d=\pi$ for simplicity. It is seen that the solid line in
Fig.~\ref{fig1} closely follows the theoretical prediction (\ref{10})
(dashed line), i.e. the decay rate of Bloch oscillations is
given by the rate $\tilde{\gamma}$ of spontaneous emission.
Simultaneously with the decay of the mean velocity, the mean
squared velocity relaxes to its equilibrium value
$v^2_{st}=v_0^2/2$, which corresponds to a uniform distribution of
the atoms over the Brillouin zone (so-called recoil heating, see
inset in Fig.~\ref{fig1}).

\begin{figure}
\center
\includegraphics[width=12.5cm, clip]{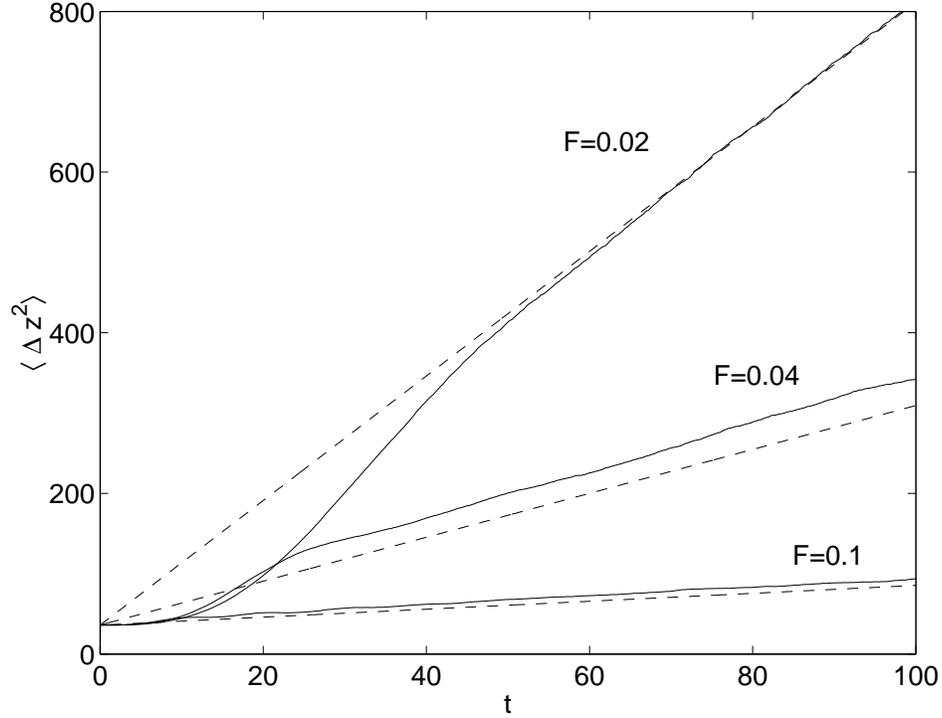}
\caption{Tight-binding model. Dispersion of the wave packet as a
function of time for $\tilde{\gamma}=0.05$. The slopes of the dashed lines 
are given by values of the diffusion coefficient (\ref{11}).} \label{fig2}
\end{figure}

Since Bloch oscillations decay after a transient time $t\sim
1/\tilde{\gamma}$ it might be naively thought that in the
stationary regime $t\gg 1/\tilde{\gamma}$ there is no difference
between the cases $F\ne0$ and $F=0$. Although this is true for the
momentum distribution, the difference appears when we analyze the
atomic dynamics in coordinate space. Indeed, according to the
``classical model" (\ref{8}) the velocity correlation function
$R_t(\tau)=\langle v(t+\tau)v(t)\rangle$ does not depend on $t$ in
the stationary regime and obeys
%************************************************************
\begin{equation}
\label{13}
R(\tau)=v^2_{st}\exp(-\tilde{\gamma} \tau)\cos(\omega_B \tau) \;.
\end{equation}
Substituting Eq.~(\ref{13}) into the equation for the mean
squared displacement,
%************************************************************
\begin{equation}
\label{14}
\langle z^2(t)\rangle=\langle \left( \int_0^t v(t'){\rm d}
t'\right)^2\rangle \approx 2t\int_0^\infty R(\tau) {\rm d}\tau \;,
\end{equation}
we obtain
%************************************************************
\begin{equation}
\label{11}
\langle z^2(t)\rangle\sim D t \;,\quad D=2v^2_{st}
\,\frac{\tilde{\gamma}}{\omega^2_B+\tilde{\gamma}^2} \;.
\end{equation}
Thus the static force suppresses diffusion caused by the recoil heating.
As an example, Fig.~\ref{fig2} shows numerical results
(obtained on the basis of the tight-binding model) for the mean
squared displacement $\langle \Delta z^2(t)\rangle=\langle
z^2(t)\rangle-\langle z(t)\rangle^2$ for some values of $F$
(solid lines) together with the diffusion law
$\langle \Delta z^2(t)\rangle=Dt$
(dashed lines) where the diffusion coefficient $D$
is obtained from (\ref{11}). Concluding this section we stress that
Eq.~(\ref{11}) refers only to the stationary regime $t\gg
1/\tilde{\gamma}$. In the opposite limit the atomic motion is
essentially oscillatory (large static force) or ballistic
(weak force) and $\langle \Delta z^2(t)\rangle\sim t^2$.

%%%%%%%%%%%%%%%%%%%%%%%%%%%%%%%%%%%%%%%%%%%%%%%%%%%%%%%%%%%
\section{Decoherence by recoil heating}

\begin{figure}
\center
\includegraphics[width=12.5cm, clip]{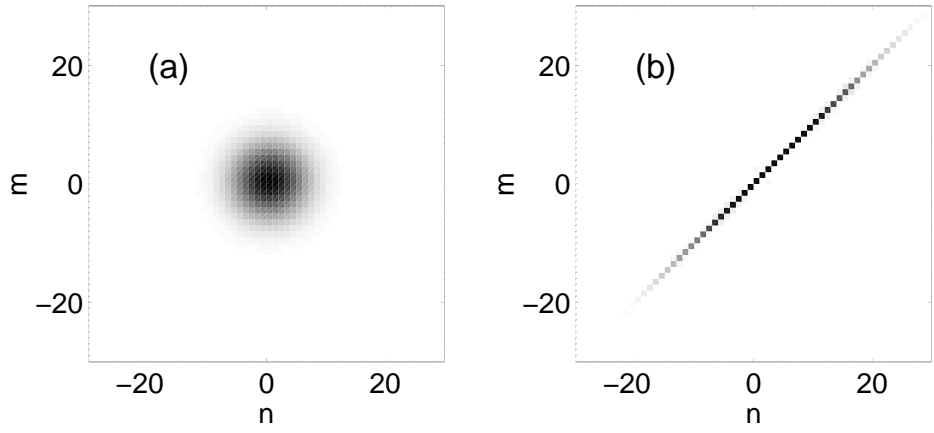}
\caption{Absolute values of the density matrix elements $\rho_{n,m}(t)$
as a gray-scaled map for $t=0$ (a) and $t=5T_B$ (b). The parameters
are the same as in Fig. 1.} \label{fig8}
\end{figure}

The results reported in the previous section can be also viewed
as a decoherence process. Indeed, in the single-band approximation
the dynamics of the system is characterized by the density matrix
$\rho(t)=\sum_{n,m} \rho_{n,m}(t) |m \rangle\langle n|$. Substituting
Eqs.~(\ref{6})-(\ref{7}) into the master equation (\ref{1})
and taking into account that $P(u)\approx1/2$ we obtain
%**********************************************************
\begin{equation}
\label{20}
\frac{{\rm d}\rho_{n,m}(t)}{{\rm d} t}\approx-\frac{i\Delta}{2\hbar}
\sum_{+,-}(\rho_{n\pm 1,m}-\rho_{n,m\pm 1})-
i\omega_B(n-m)\rho_{n,m}-\tilde{\gamma}(1-\delta_{n,m})\rho_{n,m} \;.
\end{equation}
The last term in this equation causes the decay of off-diagonal matrix
elements. Thus the density matrix of the system tends to be diagonal in the
basis of the localized Wannier states. As an illustration to this
statement Fig.~\ref{fig8} shows the absolute values of $\rho_{n,m}(t)$
as a gray-scaled map for the parameters of Fig.~\ref{fig1} and
$t=0$ and $t=5T_B$. It is seen in the figure that at
$t>1/\tilde{\gamma}$ only the diagonal and nearest to diagonal
matrix elements $\rho_{n,n+1}=\rho^*_{n+1,n}$
have non-negligible values. This property of the ''stationary"
density matrix allows us to obtain Eq.~(\ref{11}) of the
previous section without appealing to the classical model
(\ref{8})-(\ref{12}). The derivation is as follows.

First, we estimate the off-diagonal matrix elements $\rho_{n+1,n}$.
The formal solution for these elements reads
%**********************************************************
\begin{equation}
\label{21}
\rho_{n+1,n}(t)=\frac{i\Delta}{2\hbar}
\exp(-i\omega_B t-\tilde{\gamma} t) \int_0^t {\rm d} t'
(\rho_{n+1,n+1}-\rho_{n,n})
\exp(i\omega_B t'+\tilde{\gamma} t')
\end{equation}
(here we neglect all off-diagonal elements except the ones nearest to the
main diagonal). Because in the stationary regime the characteristic rate
of change of the diagonal matrix elements is much smaller than $\omega_B$,
we obtain
%**********************************************************
\begin{equation}
\label{22}
\rho_{n+1,n}(t)\approx\frac{i\Delta}{2\hbar}
\frac{\rho_{n+1,n+1}-\rho_{n,n}}{i\omega_B+\tilde{\gamma}} \;.
\end{equation}
Next, substituting this estimate into the equation for the diagonal
matrix elements yields the rate equation
%**********************************************************
\begin{equation}
\label{23}
\frac{{\rm d}\rho_{n,n}(t)}{{\rm d} t}=\left(\frac{\Delta}{2\hbar}\right)^2
\frac{2\tilde{\gamma}}{\omega_B^2+\tilde{\gamma}^2}\,
(\rho_{n+1,n+1}-2\rho_{n,n}+\rho_{n-1,n-1}) \;.
\end{equation}
Finally, approximating $(\rho_{n+1,n+1}-2\rho_{n,n}+\rho_{n-1,n-1})/d^2$
by $\partial^2\rho/\partial z^2$ we end up with the diffusion equation
%**********************************************************
\begin{equation}
\label{24}
\frac{\partial\rho}{\partial t}=\frac{D}{2}
\frac{\partial^2\rho}{\partial z^2} \;,
\end{equation}
with the same diffusion coefficient $D$ as given in Eq.~(\ref{11}).

To summarize, the recoil heating reduces the coherence length of
the atomic wave function to that of the localization length of the
Wannier function which, in turn, is of the order of the
optical lattice period $\pi/k_L$. Obviously, under this condition
a direct observation of Bloch oscillations is impossible.
However, they show up indirectly as a ''correction" to the
diffusion coefficient $D$.

%%%%%%%%%%%%%%%%%%%%%%%%%%%%%%%%%%%%%%%%%%%%%%%%%%%%%%%
\section{Beyond the single-band results}

\begin{figure}
\center
\includegraphics[width=12.5cm, clip]{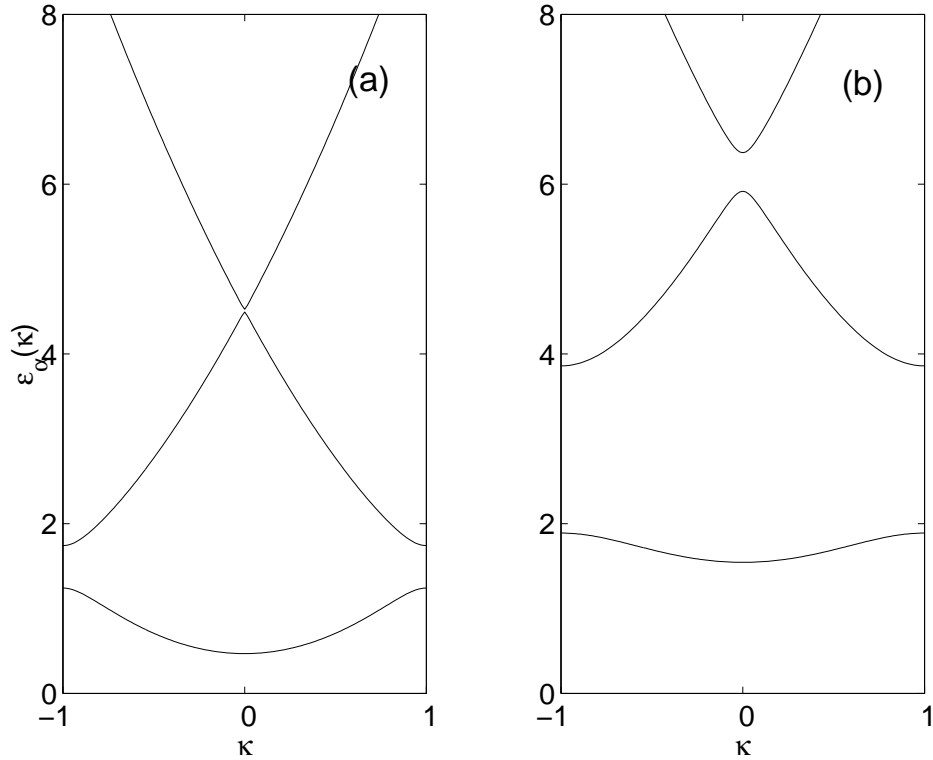}
\caption{Bloch band spectrum of the system (\ref{15}) for $F=0$
and $U=1$ (a) and $U=4$ (b).} \label{fig3}
\end{figure}

We proceed with the analysis of Bloch oscillation on the basis of
Eqs.~(\ref{1}) -- (\ref{3}), i.e. beyond the tight-binding
(single-band) approximation. It is convenient to use scaled
variables, where the length is measured in units of the laser
wave-length ($z\rightarrow k_L z$), the energy in units of the
recoil energy $E_R=\hbar^2 k_L^2/2M$, and the time is scaled on
the basis of the recoil frequency ($t\rightarrow E_R t/\hbar$).
Then the Hamiltonian (\ref{2}) takes the form
%************************************************************
\begin{equation}
\label{15}
H=p^2-U\cos^2(z)+Fz \;,
\end{equation}
where $p=-i{\rm d}/{\rm d}z$, $U=(\hbar\Omega^2/\delta)/E_R$, and
$F$ is the scaled static force. In what follows we restrict
ourselves by considering two different values for amplitude of
the optical potential: $U=1$, which we shall refer to as case (a),
and case (b), where $U=4$. The band spectrum of the system in
these two cases is depicted in Fig.~\ref{fig3}. It is seen in the
figure that for $U=1$ there is an essential deviation of the
ground-Bloch-band dispersion relation from the cosine law, while for
$U=4$ the relation $\epsilon_0(\kappa)=-\Delta\cos(\pi\kappa)$
holds with good accuracy. The more important difference of two
spectra, however, is the size of the energy gap between the
ground and ``first excited" bands. Indeed, according to the
Landau-Zener theory this gap defines the probability of interband
tunneling, which decreases exponentially as a function of squared
energy gap \cite{Land32}. Thus, in the case (b) the rate of
Landau-Zener tunneling is negligible in comparison with case (a).

\begin{figure}
\center
\includegraphics[width=12.5cm, clip]{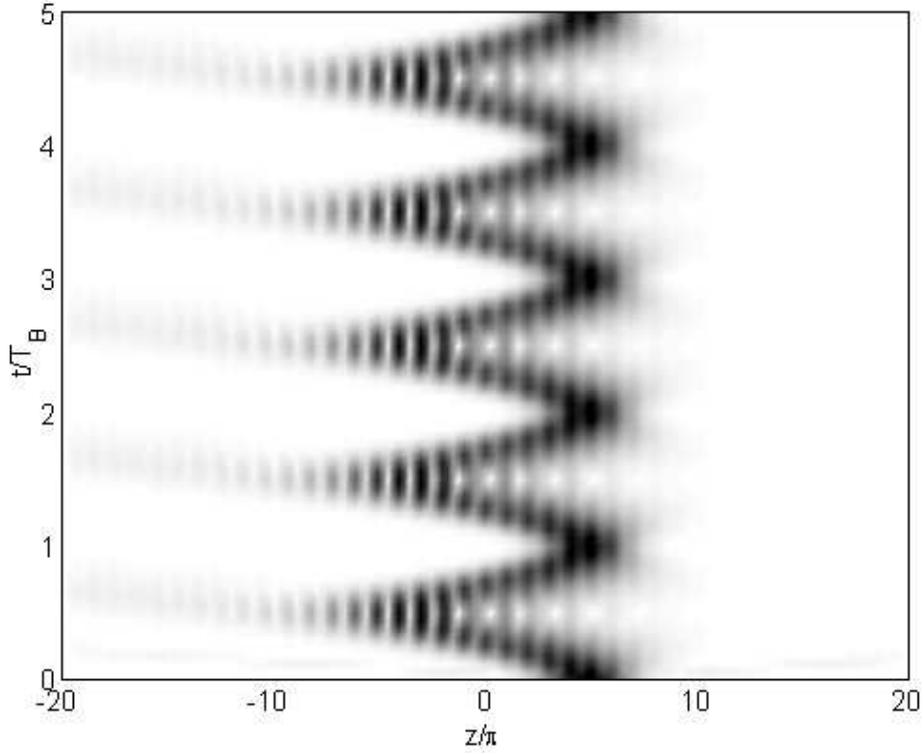}
\caption{Coherent (undamped) Bloch oscillations. Absolute value of
the atomic wave function is shown as a gray-scaled map. The
system parameters are $U=1$ and $F=0.025$.} \label{fig4}
\end{figure}

\begin{figure}
\center
\includegraphics[width=12.5cm, clip]{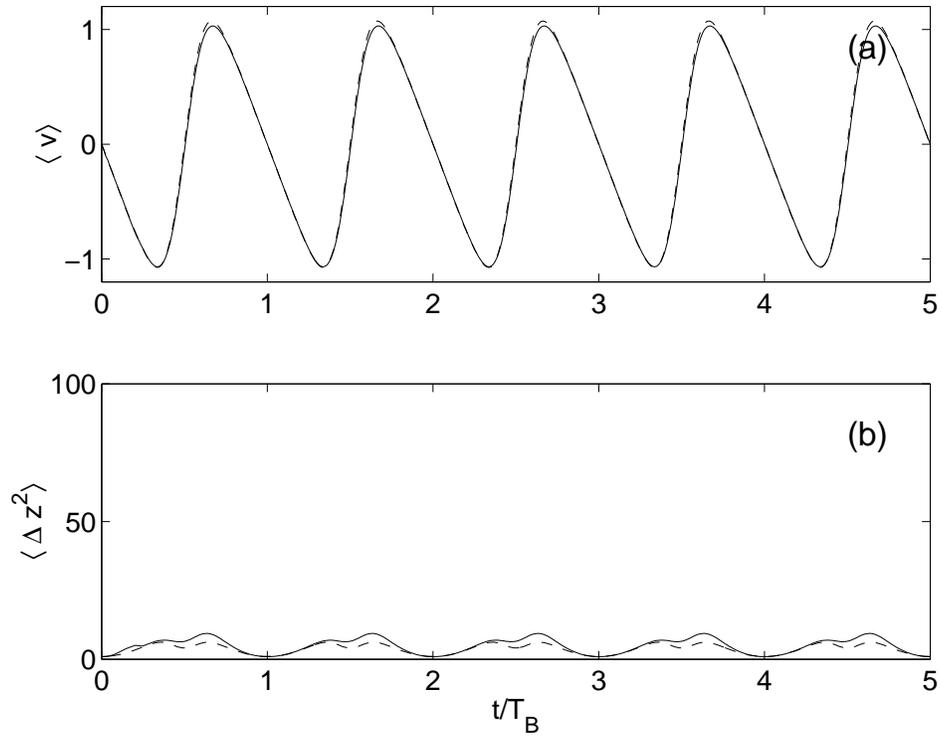}
\caption{Undamped Bloch oscillations. The mean atomic velocity (a)
and the dispersion of the wave packet (b) are shown as the
functions of time. Parameters are the same as in Fig.~4. The
dashed lines show the results of the single-band
approximation.} \label{fig5}
\end{figure}

Figure \ref{fig4} shows the dynamics of an atomic wave packet for
$\tilde{\gamma}=0$, $U=1$, and $F=0.025$. As expected the atoms
perform a periodic oscillations with Bloch period $T_B=2/F$. It
is also seen in the figure that during each Bloch cycle (when
the wavepacket is reflected at its leftmost position) a fraction
of probability is ``emitted"
in the negative direction. In terms of the Bloch-band spectrum this
effect corresponds to Landau-Zener tunneling between the ground
and upper bands, discussed above. Experimentally, this phenomenon
was observed in Ref.~\cite{Ande98} and its complete theoretical
description is given in Ref.~\cite{PR02} by using the formalism of
resonance (metastable) Wannier-Stark states. Because of
tunneling the probability $P(t)$ to find the atom in any finite
interval (larger than the amplitude of Bloch oscillations)
exponentially decreases with time. In what follows, we quantify
this process by the increment $\nu$, which we refer to as the
depletion constant \cite{remark}. In numerical simulation the
depletion constant was found by approximating the function
%************************************************************
\begin{equation}
\label{16}
P(t)=\int_{-20\pi}^{20\pi}{\rm d}z |\psi(z,t)|^2
\end{equation}
by an exponential function. This gives $\nu=2.5\cdot10^{-4}$ and
$\nu<10^{-6}$ for the cases (a) and (b), respectively. 
%which should be compared with the exact values $\nu=?$ and $\nu=?$,
%found by using the method of Ref.~\cite{PR02}. 
For the sake of future reference the solid lines in Fig.~\ref{fig5} 
show the dynamics of the mean atomic velocity $\langle v(t)\rangle$,
calculated as
%************************************************************
\begin{equation}
\label{17}
\langle v(t)\rangle=\int_{-20\pi}^{20\pi}{\rm d}z \psi^*(z,t)
\left(-i2\frac{\partial}{\partial z}\right) \psi(z,t) \;,
\end{equation}
and the wave packet dispersion $\langle \Delta z^2(t)\rangle= \langle
z^2(t)\rangle - \langle z(t)\rangle^2$. To compensate the
decrease of probability, both $\langle v(t)\rangle$ and $\langle
\Delta z^2(t)\rangle$ are normalized by dividing by $P(t)$.
The asymmetry of the velocity oscillations obviously reflects the
deviation of the actual dispersion relation $\epsilon_0(\kappa)$
from the cosine dispersion relation. The dashed lines in the
figure correspond to the tight-binding approximation of Sec.~3
where, to take into account the asymmetry of the oscillations,
the tri-diagonal matrix (\ref{6}) is substituted by few-diagonal
matrix with off-diagonal matrix elements given by the
coefficients of Fourier transform of $\epsilon_0(\kappa)$. A
reasonable coincidence is noticed.

We proceed with the case of nonzero rate of the spontaneous
emission. First we study the dependence of the depletion constant
$\nu$ on $\tilde{\gamma}$. These studies are summarized in
Table \ref{tab1}. It is seen from the table that the spontaneous emission
strongly enhances the ``tunneling" decay of the system. This can
be understood by noting that the recoil operator may ``kick out"
the atom from the ground Bloch band. (More formally, given
$\psi(z)$ belonging to the subspace of Hilbert space spanned by
Bloch waves with zero band index, $L_u\psi(z)$ generally does not
belong to this subspace.) Because the rate of Landau-Zener
tunneling increases with the band index, this causes a faster
decay of $P(t)$.

\begin{figure}
\center
\includegraphics[width=12.5cm, clip]{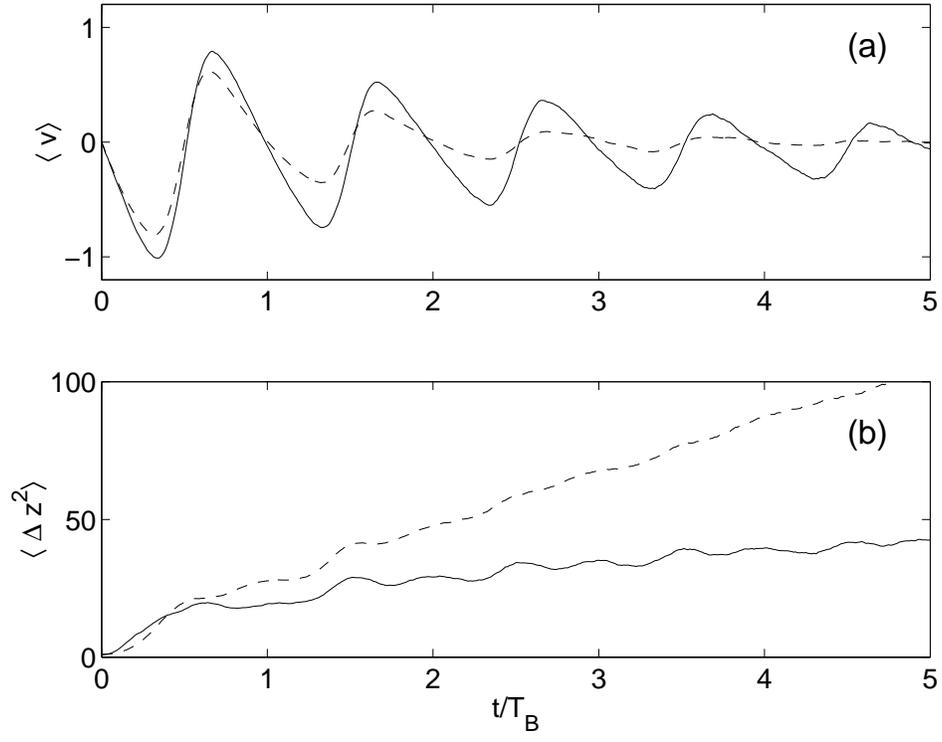}
\caption{Damped Bloch oscillations. The same as in Fig.~5 but for
$\tilde{\gamma}=0.01$.} \label{fig6}
\end{figure}

Since the depletion of the ground band introduces an additional decay
mechanism, it might be naively expected that the decay of Bloch
oscillations should be faster than it is predicted by the
tight-binding model. However, the real situation appears to be
inverse. This is illustrated in Fig.~\ref{fig6}, where the dashed
lines are predictions of the tight-binding model and the solid
lines show actual behaviour of the system. (The system parameters
are the same as in Fig.~\ref{fig5} but $\tilde{\gamma}=0.01$.)
The explanation for this effect is given below. As follows from
the explicit form of the recoil operator (\ref{3}), it multiplies
the Bloch wave by a plane wave with a wave vector varying from zero
to that of the size of the Brillouin zone. Within the single band
approximation this always results in a change of the quasimomentum
$\kappa$ of Bloch wave by the same value. In the reality, however,
only small values of the wave-vector (``weak kicks") cause a
simple change of the quasimomentum. The large values (``strong
kicks"), as explained above, may remove the system from the
ground band. As a result, the redistribution of the quasimomentum
$\kappa$ over the entire Brillouin zone is slowed down and the oscillations
of the mean velocity (normalized by survival probability $P(t)$)
decays slower.

\begin{figure}
\center
\includegraphics[width=12.5cm, clip]{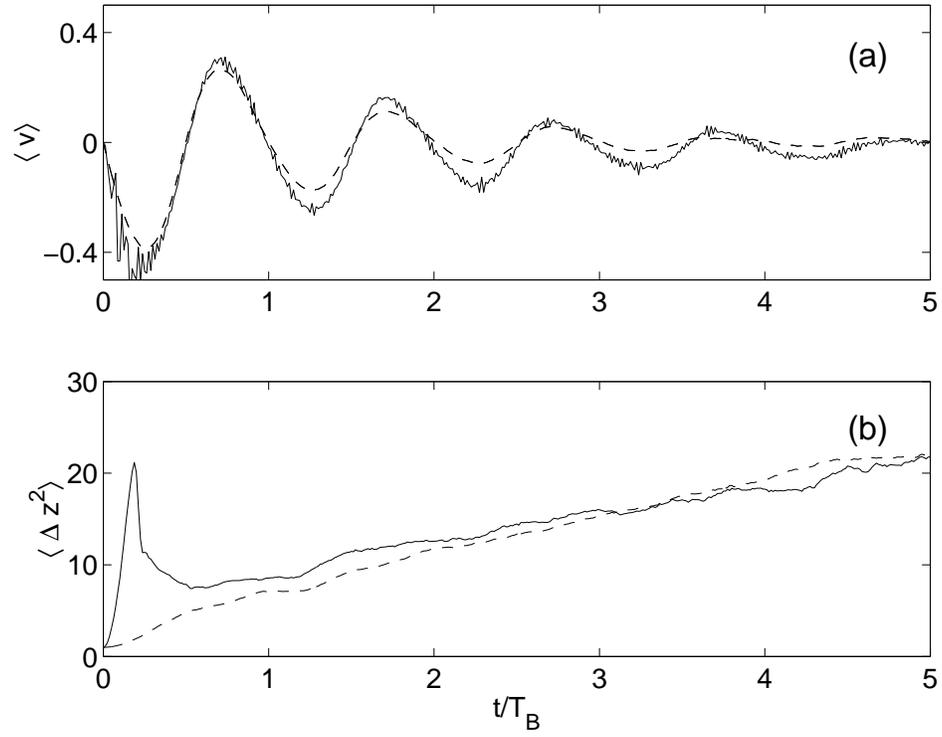}
\caption{Damped Bloch oscillations. The same as in Fig.~6 but for
$U=4$. (Weak oscillations of the velocity is an artificial fact} \label{fig7}
\end{figure}

The results of a numerical simulation in the case $U=4$ are shown
in Fig.~\ref{fig7}. Peak-like behaviour of the dispersion at short times
time origin is a short-time transient phenomenon depending on the
initial condition. (In principle, this peak can be removed by an
appropriate choice of the initial wave packet.) Ignoring this
transient phenomenon, the overall dynamics of the system is now
closer to that predicted by the tight-binding model. This is
actually not surprising, because the tight-binding
approximation is more reliable for larger values of the energy gap.

%%%%%%%%%%%%%%%%%%%%%%%%%%%%%%%%%%%%%%%%%%%%%%%%%%%%%%%%%%%%%%%%
\section{Summary}

We analyzed Bloch oscillations of neutral atoms in the presence
of spontaneous emission. It is shown that random recoil kicks
(coming from spontaneously emitted photons) cause a decay of
Bloch oscillations through the dephasing of Bloch waves. The
decay of the mean atomic velocity is accompanied by a diffusive
spreading of the atoms in configuration space. We studied both
of these processes by using the tight-binding approximation and by
direct numerical simulation of the wave packet dynamics.

Within the tight-binding model, the increment of Bloch oscillations
decay is proven to coincide with the rate of spontaneous emission
$\tilde{\gamma}$ which, in turn, is given by the product of the
excitation probability of the upper level and its natural width.
The diffusion coefficient depends additionally on the value of the
Bloch frequency $\omega_B$ (which is proportional to magnitude of
the static force) and is smaller for larger values of the Bloch
frequency. Thus, the static force suppresses the spreading of the atoms
caused by the recoil heating.

Direct numerical simulation of the system dynamics confirms
qualitatively these results of the tight-binding model. However,
because of the failure of the single-band approximation, the real
situation appears to be more complicated. It is shown that the
recoil kicks depletes the ground Bloch band with a rate
depending on the (scaled) spontaneous emission rate $\tilde{\gamma}$ and
the size of the energy gap following the ground band.
Surprisingly, the depletion of the Bloch
bands affects the decay of Bloch oscillations in a
counter-intuitive way: it {\it slows down\/} the decay of the
oscillations.

%%%%%%%%%%%%%%%%%%%%%%%%%%%%%%%%%%%%%%%%%%%%%%%%%%%%%%%%%%%%%%%%%
%\bibliographystyle{aps}
%\bibliography{abbrev,paper90,emission,publist}

\begin{thebibliography}{10}

\bibitem{Bloc28}
F.~Bloch, Z. Phys \textbf{52}, 555 (1928).

\bibitem{Feld92}
J.~Feldmann {\it et. al.}, Phys. Rev. B \textbf{46}, 7252 (1992).

\bibitem{Leo92}
K.~Leo {\it et. al.}, Solid State Comm. \textbf{84}, 943 (1992).

\bibitem{Daha96}
M.~BenDahan {\it et. al.}, Phys. Rev. Lett. \textbf{76}, 4508 (1996).

\bibitem{Raiz97}
M.~G. Raizen, C.~Salomon, and Qian Niu, Physics Today
\textbf{{\rm July}}, 30 (1997).

\bibitem{Ande98}
B.~P. Anderson and M.~A. Kasevich, Science \textbf{282}, 1686 (1998).

\bibitem{PLA00}
M.~Gl{\"u}ck, A.~R. Kolovsky, and H.~J. Korsch, Phys. Lett. A
\textbf{276}, 167 (2000).

\bibitem{Mors01}
O.~Morsch {\it et. al.}, Phys. Rev. Lett. \textbf{87}, 140402 (2001).

\bibitem{Thom02}
Q.~Thommen, J.~C. Garreau, and V.~Zehnle, Phys. Rev. A \textbf{65},
053406 (2002).

\bibitem{PR02}
M.~Gl{\"u}ck, A.~R. Kolovsky, and H.~J. Korsch, Phys. Rep., in press (2002).

\bibitem{Leto81}
V.~S. Letokhov and V.~G. Minogin, Phys. Rep. \textbf{73}, 1 (1981).

\bibitem{Adam94}
C.~S. Adams, M.~Sigel, and J.~Mlynek, Phys. Rep. \textbf{240}, 143 (1994).

\bibitem{Grah96}
R.~Graham and S.Miyazaki, Phys. Rev. A \textbf{53}, 2683 (1996).

\bibitem{Goet96}
P.~Goetsch and R.~Graham, Phys. Rev. A \textbf{54}, 5345 (1996).

\bibitem{Meystre}
P.~Meystre and M.~Sargent III, \emph{Elements of quantum optics},
Springer, 1999.

\bibitem{Land32}
L.~D. Landau, Z. Sov. \textbf{1}, 46 (1932).

\bibitem{Zene34}
C.~Zener, Proc. R. Soc. A \textbf{145}, 523 (1934).

\bibitem{Bhar97}
C.~F. Bharucha {\it et. al.}, Phys. Rev. A \textbf{55}, R857 (1997).

\bibitem{JOPB00a}
M.~Gl{\"u}ck, A.~R. Kolovsky, and H.~J. Korsch, Journal of Optics B: Quantum
and Semiclassical Optics \textbf{2}, 694 (2000).

%\bibitem{remark2}
%For $\tilde{\gamma}=0$ the conditional probability (\ref{12})
%obviously reduces to $W(p,p',t)=\delta(p-p'-Ft)$.

\bibitem{remark}
Alternatively, the discussed probability can be considered as an
occupation probability of the ground Bloch band. For
$\tilde{\gamma}\approx0$ the depletion constant was measured in
\cite{Bhar97}.

\end{thebibliography}

%%%%%%%%%%%%%%%%%%%%%%%%%%%%%%%%%%%%%%%%%%%%%%%%%%%%%%%%%%%%%%%
\begin{table}[b]
\caption{Depletion constant $\nu$ for different rate of
spontaneous emission $\tilde{\gamma}$. The value of the static
force is $F=0.025$.}
\label{tab1}
\begin{ruledtabular}
\begin{tabular}{ccccc}
$\tilde{\gamma}$&0.0&0.001&0.01&0.05\\
\colrule
$U=1$&$2.5\cdot10^{-4}$&$6.0\cdot10^{-4}$&$3.0\cdot10^{-3}$&$1.0\cdot10^{-2}$\\
$U=4$&      $< 10^{-6}$&$1.5\cdot10^{-4}$&$1.5\cdot10^{-3}$&$0.9\cdot10^{-2}$
\end{tabular}
\end{ruledtabular}
\end{table}

%%%%%%%%%%%%%%%%%%%%%%%%%%%%%%%%%%%%%%%%%%%%%%%%%%%%%%%%%%%%%%%%%%
\newpage

\end{document}